\documentclass[10pt]{amsart}
\usepackage{amsmath}
\usepackage{amsfonts}
\usepackage{amssymb}
\usepackage{epsfig}
\usepackage{graphicx}

\title[Geometric Frustration]{Geometric Frustration and Interparticle Gap Size Distributions in Ordered Hexagonal Polydisperse Disk Packs}
\date{November 5, 2007}
\author[Daniel P. Snowman]{Daniel P. Snowman}
\address{Department of Physical Sciences, Rhode Island College, Providence, RI 02908}
\email{dsnowman@ric.edu}

\begin{document}

\begin{abstract}
This work analyzes the distribution and size of interparticle gaps arising in an ensemble of hexagonal unit structures in the xy plane when packing disks with a Gaussian distribution of radii with mean  (r) and standard deviation $\Delta r$.  During the course of this investigation an equivalency is established between gaps arising in hexagonal unit structure packs and nine-ball billiard rack patterns.  An analytic expression is derived for the probability distribution and location of interparticle gaps of magnitude $\Gamma$.  Due to the number of variables and large number of possible arrangements, a Monte Carlo simulation has been conducted to complement and probe the analytic form for three very different systems:  i) billiard balls with Billiard Congress of America (BCA) specifications, ii) US pennies with specifications of the US Mint, and iii) a hypothetical system with $r = 1.0 m$ and $\Delta r = 1x10^{-10}$ m corresponding to the scale of one atomic radius.  In each case, probability density distributions of gap sizes have been calculated for those $\Delta r$ above, and also for 2$\Delta r$ and 0.5$\Delta r$, respectively.  A general result is presented for the probability of a nonzero normalized ($\frac{\Gamma}{\Delta r}$) gap size arising, $P(\Gamma \geq \alpha \Delta r)= 1-0.124\alpha$, where $\alpha$ is a constant $\leq 5.0$.  This curious result reflects the phenomenon of geometric frustration; the inability of the system to simultaneously satisfy all geometric constraints required by a perfect-rack sans interparticle gaps.
\end{abstract}


\maketitle

\section{Introduction}

\subsection{Motivation}

This work analyzes the distribution and size of interparticle gaps arising, locally, in an ensemble of non-interacting, hexagonal unit structures in the xy plane when packing disks with a Gaussian distribution of radii.  During the course of this investigation an equivalency is established between gaps arising in hexagonal unit structure packs and nine-ball billiard rack patterns, see Figure 1.  In fact, the motivation behind this work was the curious difficulty in preparing a gap-free rack of nine-ball billiards.  This study maintains that upon packing disks with a Gaussian variation in radii, a perfect-pack, hexagonal and planar, is an illusion and is not attainable in practice due to geometric frustration. A perfect rack being one in which each disk contacts each of its neighbors with no gaps present.  This paper explores this curious observable phenomenon for the simplified case of a 'nine-ball rack' of disks with radii that vary according to a Gaussian distribution with standard deviation, $\sigma=\Delta r$.  Variation in disk radii ($\Delta r$) arising naturally due to usage and/or limitations in mill tolerance.  

Due to the initial motivation for this study, gaps arising due to local variables (i.e. random variation in disk radii) have been the primary focus of this study.  Thus, the term, ensemble, refers to a large number of non-interacting unit structures as shown in Figure 4.

Intriguing parallels exist between this investigation and other physical systems with frustration present.  Frustration in this context refers to the inability of the system to simultaneously satisfy all constraints present.  Typically, in physical systems, these constraints are related directly to the energy and entropy of the system.  The present study is subject to geometric frustration that arises in that three circles (see below for development), and their respective equations, must be satisfied by a single $\{x,y\}$.  Ultimately, geometric frustration arises if the system with circles of three different radii and centers do not share a common intersection point.  The center and radii of each of these circles having an element of randomness present.  Thus, this study analytically explores the nature of this geometric frustration that reveals itself as an interparticle gap in ordered, hexagonal packings of disks in the plane.  In particular, this investigation, i) derives an analytic expression for the interparticle gap of magnitude $\Gamma$ with an associated probability distribution, ii) explores all physically realizable configurations of hexagonal disk packs, detailing resulting interparticle gap locations (see Figure 5), and, iii) performs a Monte Carlo simulation to probe the resulting probability density distribution of interparticle gaps arising in an ensemble of ${10}^5$ units.

\subsection{Disk Packings}

Physicists, chemists, engineers, geologists, biologists and mathematicians have found many applications of hard disk packings as simiplified models for the study of granular materials ~\cite{Bideau}, random media ~\cite{Torquato1}, glasses ~\cite{Zallen}, liquids ~\cite{Hansen}, amorphous solids ~\cite{Zallen}, cell pattern formation, investigation of island formation in metallic films ~\cite{Blackman}, segregation problems ~\cite{Troadec}, turbulence ~\cite{Manna} and plate tectonics ~\cite{Herrmann}.  The reader is directed to Aste and Weaire ~\cite{Aste2} for a quality account of many challenging theoretical problems and applications related to the field of hard disk packings.  

An earlier study by Cowan ~\cite{Cowan1} employed a random packing strategy, following the Renyi parking scheme, as equal-sized disks were placed along the circumference of a central disk.  This investigation probabilistically probed the number of gaps arising when the central disk was in contact with K neighbors.  Cowan`s study, much like the work presented in this paper, limited its scope to the immediate neighborhood of any one disk of the larger ensemble.  Cowan later ~\cite{Cowan2} considered the entire ensemble and established bounds on the density of the disk packing.  The study presented in this paper differs as we consider ordered packing of disks with a Gaussian distribution of radii.  The randomness lies in the disk-size, not the packing strategy.  
Aste ~\cite{Aste1} considered disordered packing of disks with random sizes and positions with the primary result being a power-law distribution for the size distribution of disks that densely cover the plane.  Most intriguing in this study is the agreement found with an experimental investigation involving the vapor deposition of tin drops onto a hot substrate resulting in breath figures.  

Many of the more recent studies in the field of hard disk or sphere packing have dealt with the phenomenon of jamming ~\cite{Torquato2},~\cite{Donev3}.  In most cases, these packings consider packing fractions ~\cite{Chaikin}, density fluctuations ~\cite{Donev1} and possible glass transitions ~\cite{Donev2} for various classification of jammings ~\cite{Torquato2}, for a large number of particles. Our study does not directly investigate jamming as we consider a large number of noninteracting unit structures of ordered, hexagonal disk packs with a random distribution of disk radii.  The results of the present study, however, may find direct application to these works as the distribution of interparticle gap sizes in these unit structures directly affect the response of granular packings when subjected to a disturbing force. ~\nocite{Dubejko}, ~\nocite{Kuperberg}, ~\nocite{Lubachevsky} 

\subsection{Circle Packings}

A conjecture by Thurston ~\cite{Thurston} proposed an intimate connection between circle-packings and analytic function theory; that is, the rigidity associated with analytic functions also exists with circle-packings.  Rodin and Sullivan proved the corresponding convergence ~\cite{Rodin}.  Many other profound results have been forthcoming since and the interested reader is directed to Stephenson for a comprehensive ~\cite{Stephenson2} review of the field, or, see ~\cite{Stephenson1} for a short introduction to the topic.   

It is important to note that a distinction exists between a circle packing and sphere packing in two dimensions.  Circle packings have an underlying pattern of tangencies, essentially consisting of an abstract graph, with a collection of vertices and connecting edges, that ultimately determines the circle packing pattern that arises as each circle adjusts its radius until the underlying pattern of tangencies is satisfied.

The work presented in the present paper, technically, is not a circle packing since an underlying pattern of tangencies was not used to generate the packing.  The reader should note, however, that after the ordered packing of randomly sized circles (or disks) is complete, a pattern of tangencies could be prescribed ex post facto.  Thus, this study could be considered a pseudo-circle-packing investigation with randomness introduced in the underlying pattern of tangencies.  This randomness would affect both features of the underlying associated graph:  the length of each edge and the location of each node.  

\section{Packing Strategy}
\subsection{Nine-ball racks and packing order}

Figure 1 shows a classic nine-ball diamond pattern rack of billiard balls.  Only two balls (the	 one-ball and nine-ball, respectively) have predetermined positions in a proper nine-ball billiard rack; the one-ball at the head of the rack and the nine-ball at the center.  The remaining seven balls, with enumerations 2-8, are placed in any of the remaining positions.  It is interesting to note that a proper nine-ball rack can be prepared in 7! unique ways for one set of radii.  Each unique configuration of disks potentially producing a unique gap size.  

\begin{figure}
\begin{center}
\leavevmode
\includegraphics[scale=0.5]{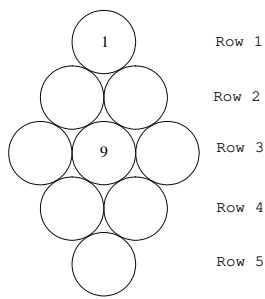} 
\end{center}
\caption{A traditional nine-ball rack pattern.}
\label{fig:Figure 1}
\end{figure}

The billiard game of nine-ball demands the lowest numbered ball on the table always be struck first. Therefore, the one-ball would be located at the top of rack and typically is the first to be positioned.  This assumes, of course, that the initial impact from the cue ball would be coming from the top of the page.  After the one-ball is placed, a row of two-balls (with any enumeration 2-8) are placed snuggly behind the one-ball.  A row of three balls with the nine-ball at the center is placed leaving no gaps.  The placement of the fourth row of balls inevitably results in an interparticle gap arising in any one of the three positions shown in Figure 2.  Finally, the remaining ball is placed.  Our study involving disk packs in this same diamond shape will involve a slightly different packing strategy.  
 
\begin{figure}
\begin{center}
\leavevmode
\includegraphics[scale=0.5]{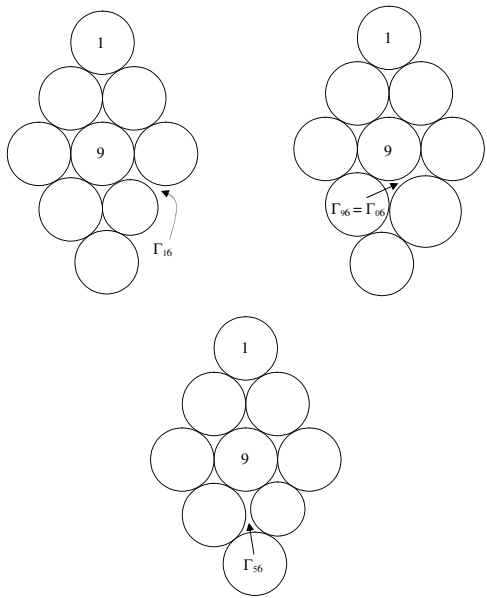} 
\end{center}
\caption{Most common interparticle gap locations in a nine-ball rack}
\label{fig:Figure 2}
\end{figure}

\subsection{Disk-packs and packing order}

This study seeks to characterize the resulting gap defects arising, thus, without any loss of generality we eliminate the disks at the top and bottom of the nine-ball pack shown in Figure 1, since neither contributes to gaps arising in our particular study.  We are now left with a hexagonal structure as shown in Figure 3a and 3b.  In our study, for computational ease, the nine-ball is placed first (at the origin) and from hereon is considered the center ball at the zeroth site as we transition away from the language of nine-ball billiards and into the arena of disk packing.

\subsection{Packing Strategy}

After the center disk (or nine ball) is placed at the origin, the surrounding disks are packed in a counter-clockwise manner beginning at site $s_1$, as shown in Figure 3.  The disk at site $s_1$ lies directly on the x-axis a distance $r_1+r_0$ from the origin.  The disk at site $s_2$ is placed so that it is in contact with the disks at sites $s_0$ and $s_1$.  Each pair of disks in contact share a common tangent.  The disk at site $s_3$ is placed so that it is in contact with the disks at sites $s_0$ and $s_2$.  This process is repeated for the disks to be placed at sites $s_4$ and $s_5$.  It is assumed that each disk to be placed does not disturb any disk previously placed.  If the seven disks have identical radii, as shown in Figure 3b, a perfect rack will result with 12 points of contact after the sixth and final disk has been placed.  However, with disk of varying radii the final disk to be placed poses potential problems since the radii are not identical and it may not be possible to satisfy all geometric constraints present, given by Equations 5-7.  

\begin{figure}
\begin{center}
\leavevmode
\includegraphics[scale=0.5]{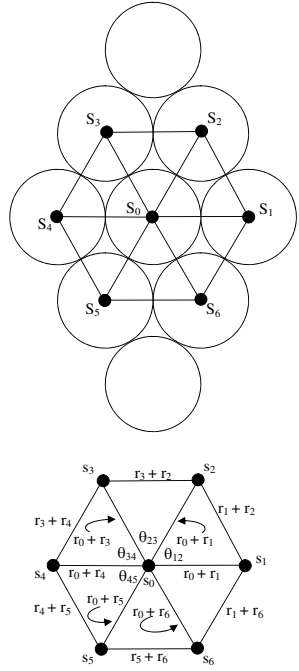} 
\end{center}
\caption{Nine-ball rack pattern and equivalent hexagonal unit structure used in this study.}
\label{fig:Figure 3}
\end{figure}

\section{Determination of Disk Location}

The location of each disk`s center is determined by its radius, disk $0$, and the disk placed immediately prior.  The Law of Cosines allows for the determination of the angle that the $i^{th}$ disk`s center will lie upon at a distance $r_0 + r_i$ for each disk $i$.   Figure 4b is drawn using the fact that each pair of disks in contact will have their two centers separated by a straight line of length $r_i + r_j$.

\begin{equation}
 \theta_{ij}=\arccos[\frac{(r_i+r_j)^2-(r_i+r_0)^2-(r_j+r_0)^2}{-2(r_i+r_0)(r_j+r_0)}]
\end{equation}

where $i$ and $j$ are immediate neighbors along the perimeter.  Using this expression and Figure 3, we can write  

\begin{equation}
 \theta_{15}=\theta_{12}+\theta_{23}+\theta_{34}+\theta_{45}
\end{equation}

Thus, in polar coordinates the center of disk 5 is located a distance $r_5 + r_0$ from the origin at an angle  $\theta_{15}$.

\begin{eqnarray}
x_5=(r_5 + r_0)\cos{\theta_{15}}\\
y_5=(r_5 + r_0)\sin{\theta_{15}}	
\end{eqnarray}

The relationship derived above for $\{x_5, y_5\}$ is exact for one set of radii, $\{r_i\}$, and single unit structure.  However, a large ensemble of these unit structures, each with a unique set of radii, {$r_i$}, requires the introduction of a probability distribution; see Section 5.  

After disks have been placed at sites 1 to 5 the final task remains:  to place a disk at site six and determine analytically any resultant interparticle gaps arising.  The placement of the disk at sites 2 to 5 is determined by the center disk and the disk placed immediately prior.  The sixth and final disk is unique in that a perfect rack would have it contact three previously placed disk at sites $s_0$, $s_5$ and $s_1$.

Contact between neighboring disks implies a separation between disk centers equal to the sum of their respective radii.  Three \textit{circle-of-contact} equations can be written with knowledge of their centers and corresponding radii.  Equation 5 represents a circle-of-contact between disks at sites 0 and 6 with the center located at $\{x_0, y_0\}=\{0,0\}$ with radius $r_0 + r_6$.  Second, equation 6 represents a circle-of-contact between disks at sites 5 and 6 with center located at $\{x_5,y_5\}=\{(r_0 + r_5) \cos{\theta_{15}},(r_0+r_5) \sin{\theta_{15}}\}$ with radius $r_5 + r_6$.  Finally, equation 7 represents a circle-of-contact between disks at sites 1 and 6 with center located at $\{x_1,y_1\}=\{r_1 + r_0,0\}$ with radius $r_1 + r_6$. 

The circle-of-contact equations are used to analytically determine all possible placements of one disk relative to its neighbor that results in contact between the two.  Visually, circles-of-contact also have a strong appeal, as shown in Figure 4.  Figure 5 qualitatively depicts all types of intersections occurring between the three circle-of-contact equations.

\begin{figure}
\begin{center}
\leavevmode
\includegraphics[scale=0.5]{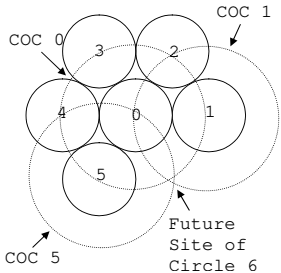} 
\end{center}
\caption{Qualitative depiction of circle-of-contact equations representing the curve along which disks at sites 0, 1 and 5, respectively; will be in contact with a disk to be placed at site 6.}
\label{fig:Figure 4}
\end{figure}

The location of the sixth disk $\{x_6, y_6\}$ must satisfy each of the three circle-of-contact equations, Equations 5-7, if it is to be in contact with all three of its neighbors at sites 0, 5, and 1.  A perfect rack will result only if these three circle-of-contact equations share a common intersection point.  If only two of the three relations intersect an interparticle gap will be present.  

\begin{eqnarray}
x^2+y^2=(r_0+r_6)^2\\
(x-x_5)^2+(y-y_5)^2=(r_5+r_6)^2\\
(x-x_1)^2+y^2=(r_1+r_6)^2
\end{eqnarray}
 		
  This study is concerned with finding those $\{x,y\}$ that satisfy at least two of the three equations above.  A triple intersection point would correspond to an $\{x,y\}$ that satisfies all three of the equations, corresponding to a perfect rack with no gap defect.  

\section{Possible Configurations and Gap Locations}

Figure 5 depicts twenty-five nontrivial, at least two of the three circle-of-contact equations share a solution, intersections in the fourth quadrant.  The fourth quadrant is the focus of this study as we attempt to place the sixth disk, see Figures 3 and 4, subject to the functional constraints of three-separate circle-of-contact equations given by Equations 5-7.  Those intersection points, shown in Figure 5, represent the location of the center of disk six resulting in contact with two of its three neighboring disks.

\begin{figure}
\begin{center}
\leavevmode
\includegraphics[scale=0.5]{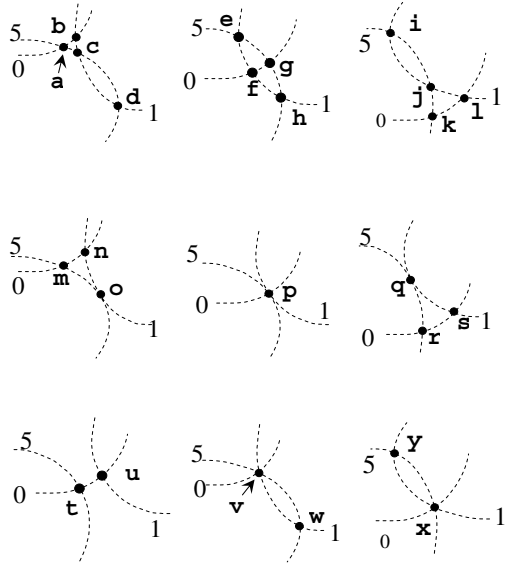} 
\end{center}
\caption{Qualitative depiction of all possible intersections of the three circle-of-contact equations.  Each intersection represents the potential, future location of disk 6}
\label{fig:Figure 5}
\end{figure}

This investigation further reduces the number of solutions considered as we eliminate those intersections not physically possible.  That is, if the location of any one of the intersection points is located inside any one of the three circle-of-contact curves, that solution is eliminated since it is not physically possible.  That is, the closest approach of any two disk centers is a sum of their respective radii.  Inspection of Figure 5 reveals that intersection points e, f, g, i, j, q, and y are not physically realizable since these intersections would demand that two neighboring disks are separated by a distance less than the sum of their respective radii.  

Of the 18 remaining solutions:  i) three triple intersection points exist (p, v, and x) corresponding to zero interparticle gap, ii) five points of intersection (a, k, m, r and t) leading to a gap between sites 5 and 6, $\Gamma_{56}$, iii) five points (b, l, n, s, and u)  of intersection leading to a gap between sites 1 and 6, $\Gamma_{16}$, and iv) five points of intersection (c, d, h, o and w) leading to a gap between sites 0 and 6, $\Gamma_{06}$.  Of these eighteen physically realizable configurations, only 13 are unique, with symmetry arguments allowing us to qualitatively equate the following; a=b, k=l, m=n, r=s and t=u.

\section{Analytic Gap Size}

After determining the disk configurations physically possible (see Section 4) for a given set of radii, the size of any interparticle gap present between disk six and any one of its neighbors is determined using Equations 8-10.

\begin{eqnarray}
\Gamma_{06}=[(x_0-x_6)^2+(y_0-y_6)^2]^{\frac{1}{2}} -(r_0+r_6)\\
\Gamma_{16}=[(x_1-x_6)^2+(y_1-y_6)^2]^{\frac{1}{2}} -(r_1+r_6)\\
\Gamma_{56}=[(x_5-x_6)^2+(y_5-y_6)^2]^{\frac{1}{2}} -(r_5+r_6)
\end{eqnarray}

Regardless of the configuration resulting, dependent upon the particular set of $r_i$, a minimum of two of the three Equations 8-10 will be zero, corresponding to those neighboring disks in contact with disk six.  The nonzero $\Gamma$ represents the interparticle gap size occurring with probability $P(\Gamma)$, shown in Equation 11.
\begin{equation} 					
P(\Gamma)={\prod_{i=0}^6} P(r_i)
\end{equation}
 with 
\begin{equation}
\sigma=\Delta r
\end{equation}
and 
\begin{equation}
P(r_i)={\frac{1}{\sqrt{2 \pi} \sigma}}Exp[\frac{-(r_i-\overline{r})^2}{2 {\sigma^2}]}
\end{equation}
	
\section{Monte Carlo Analysis}
The probability distribution for the gap size depends upon each of the seven radii, thus, due to the number of variables and the large number of possible arrangements, a numerical simulation has been conducted to complement and probe the analytic form of the gap probability distribution.  

Each simulation involves generating seven radii from a Gaussian distribution and packing them counter-clockwise starting at site 1 (as described in Section 2).  For each of $10^5$ packings, a gap size was calculated and a resultant probability density distribution produced.  This study has considered three very different disk-packings with disks of radii that follow a Gaussian distribution with mean ($r$) and standard deviation ($\Delta r$).  The systems considered were:  i) billiard balls with Billiard Congress of America (BCA) specifications, ii) US pennies with specification of the US Mint, and iii) a hypothetical system with $r = 1.0 m$ and $\Delta r = 1x10^{-10} m$ corresponding to the scale of one atomic radius.  In each case the probability density distribution of gap sizes has been calculated for those $\Delta r$ above, and also for $2\Delta r$ and $0.5 \Delta r$.

\section{Results}

In each case detailed below, it was found that the distribution of gaps arising between disk 6 and neighbors 0, 1, and 5, respectively (and computationally calculated with $\Gamma_{06}$, $\Gamma_{16}$, and $\Gamma_{56}$) are statistically indistinguishable from one another.  Thus, the analysis below pertains to gap size distributions regardless of location. 

For each of the three systems considered, the resulting probability density distributions were found to be Gaussian in nature.  Normalization of these distributions for each of these unique systems demands the integral shown in Equation 14 equal 1 over the range of positive $\Gamma$.  In each case, the best-fit Gaussians were compared to the distributions resulting from the Monte Carlo simulation via a qqplot.  Correlation coefficients were calculated for each case and showed exceptional agreement with the Gaussian fits, see Section 7.4 and Figure 14.   

\begin{equation}
\int_{0}^{\infty} \beta Exp[-\frac{\Gamma^2}{2\sigma^2}]d\Gamma=1
\end{equation}
    				
Evaluating the integral of Equation 14 leads to an expression for the normalization constant
 				
\begin{equation}
\beta=\frac{2}{\sqrt{2\pi}\sigma}
\end{equation}

Thus, each probability density distribution presented below has a Gaussian of the form shown in Equation 16.
 			
\begin{equation}
PDF(\Gamma,\sigma)=\frac{2}{\sqrt{2\pi}\sigma}Exp[-\frac{\Gamma^2}{2\sigma^2}]
\end{equation}

\subsection{ Billiard Balls}

Figure 6 represents the probability density as a function of gap size for a disk packing with disk radii and mill tolerances identical to those specified by the Billiard Congress of America (BCA);  that is, billiard balls used in competition must have radii $28.57500 \pm 0.06350 mm$.  

\begin{figure}
\begin{center}
\leavevmode
\includegraphics[scale=0.5]{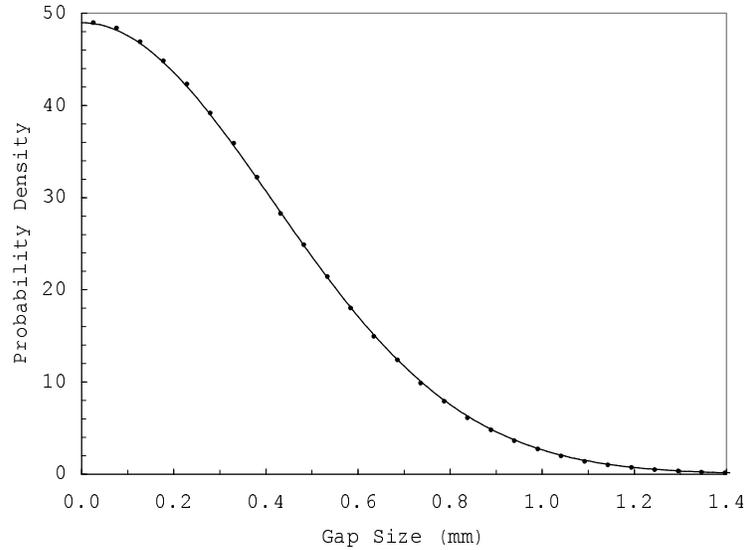} 
\end{center}
\caption{Probability density distribution for interparticle gap sizes arising in a packing of billiard balls meeting Billiard Congress of America (BCA) specifications, $r = 28.57500 mm$ and $\Delta r = 0.06350 mm$, with best-fit Gaussian. }
\label{fig:Figure 6}
\end{figure}

The probability density data shown in Figure 6 is described exceptionally well with a Gaussian best-fit with the functional form of Equation 16.  The probability density data plotted and the best-fit curve both have standard deviation $\sigma = 1.63 x 10^{-2}$ mm.  Figure 7 compares gap size probability density distributions arising using the BCA specification $\Delta r = 0.06350 mm$ to those resulting when: i) $\Delta r$ is doubled to $0.12700 mm$, and ii) $\Delta r$ is halved to $0.03175 mm$.  Doubling $\Delta r$ results in a broader distribution characterized with a standard deviation $\sigma = 3.25 x 10^{-2} mm$, consistent with both simulation data and best-fit Gaussian.  Halving $\Delta r$ results in a sharper gap size distribution characterized with a standard deviation $\sigma = 8.18 x 10^{-3} mm$, consistent with both Monte Carlo simulation data and best-fit Gaussian.  The reader should note that a doubling (halving) of $\Delta r$ results in a proportional change in the standard deviation, $\sigma$, characterizing the probability density distribution. 

\begin{figure}
\begin{center}
\leavevmode
\includegraphics[scale=0.5]{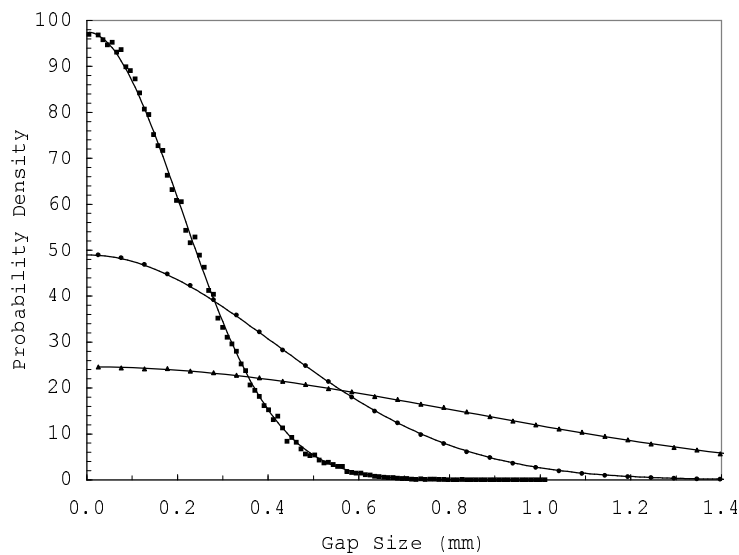} 
\end{center}
\caption{A comparison of probability density distributions for interparticle gap sizes arising in packing of billiard balls with $r = 28.5750 mm$ and i) doubled BCA spec $\Delta r = 0.12700 mm$ denoted by ($\blacktriangle$), ii) BCA spec. $\Delta r = 0.06350 mm$ denoted by ($\bullet$), and iii) halved BCA spec. to $\Delta r = 0.03157 mm$ denoted by ($\blacksquare$); with best-fit Gaussians.}
\label{fig:Figure 7}
\end{figure}

The average gap size calculated analytically using the best-fit Gaussian and Equation 17, reveal $\Gamma_{ave}= 1.29 x 10^{-2} mm$, in agreement with the Monte Carlo simulation data. 
 
\begin{equation}
\Gamma_{ave} = \int_{0}^{\infty} \Gamma \frac{2}{\sqrt{2\pi}\sigma} Exp[-\frac{\Gamma^2}{2\sigma^2}]d\Gamma
\end{equation}

\subsection{United States Pennies}

Figure 8 represents the probability density as a function of gap size for a disk packing with disk radii and mill tolerances identical to those specified by the United States Mint for the manufacture and production of US Pennies;  that is, US Pennies must have radii $9.5250 mm \pm 0.0050 mm$.  This may be the most interesting of the three cases considered in this paper since an actual, experimental disk-packing can be easily performed using seven pennies and a tabletop.  Reader beware, however, the average interparticle gaps arising is shown below to be only $2.60 x 10^{-2} mm$. 

\begin{figure}
\begin{center}
\leavevmode
\includegraphics[scale=0.5]
{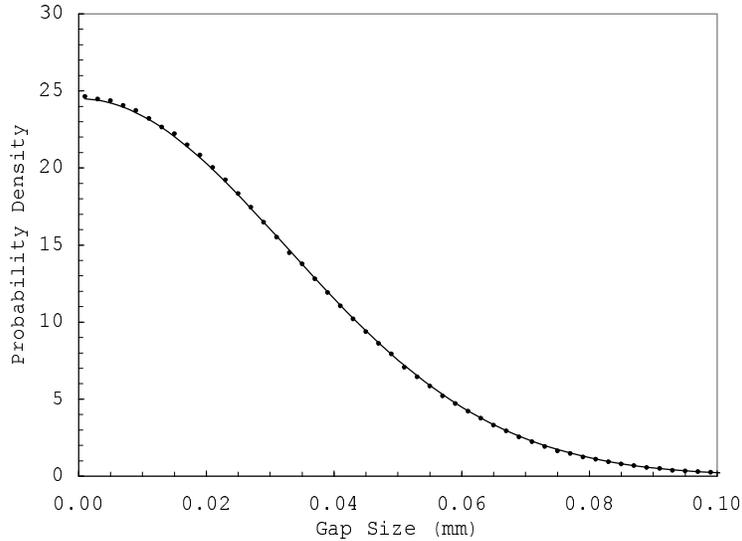} 
\end{center}
\caption{Probability density distribution for interparticle gap sizes arising in a packing of pennies with US Mint specifications $r = 9.5250 mm$ and $\Delta r = 0.0050 mm$, with best-fit Gaussian.}

\label{fig:Figure 8}
\end{figure}

The probability density distribution shown in Figure 8 hereto is represented very well with a Gaussian best-fit with the functional form of Equation 16.  The probability density data plotted and the best-fit curve both have standard deviation $\sigma = 3.26 x 10^{-2} mm$.  Figure 9 compares gap size probability density distribution arising using the US Mint specification $\Delta r=0.0050 mm$ to those resulting when: i) $\Delta r$ is doubled to $0.0100 mm$, and ii) $\Delta r$ is halved to $0.0025 mm$.  Doubling $\Delta r$ results in a broader distribution characterized with a standard deviation $\sigma = 6.45 x 10^{-2} mm$, consistent with both simulation data and best-fit Gaussian.  Halving $\Delta r$ results in a sharper distribution characterized with a standard deviation $\sigma = 1.62 x 10^{-2} mm$, consistent with both simulation data and best-fit Gaussian.  Once again, notice that doubling (halving) $\Delta r$ results in a corresponding doubling (halving) of the standard deviation, $\sigma$, the characteristic parameter defining the sharpness of the probability density distribution.

\begin{figure}
\begin{center}
\leavevmode
\includegraphics[scale=0.5]{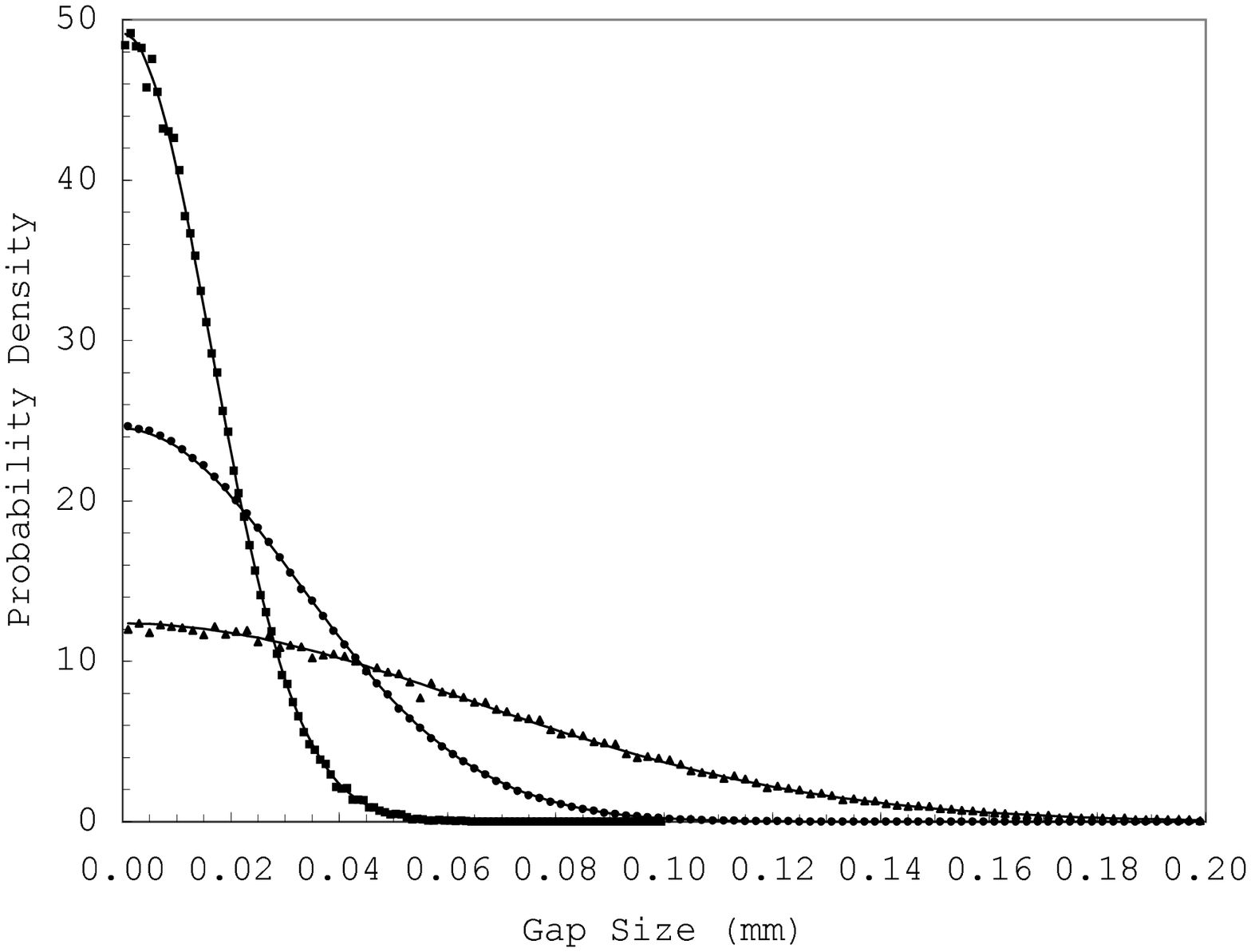} 
\end{center}
\caption{A comparison of probability density distributions for interparticle gap sizes arising in packing of United States pennies with $r = 9.525 mm$ and i) doubled US Mint spec $\Delta r = 0.0100 mm$ denoted by ($\blacktriangle$), ii) US Mint spec. $\Delta r = 0.0050 mm$ denoted by ($\bullet$), and iii) halved US Mint spec. to $\Delta r = 0.0025 mm$ denoted by ($\blacksquare$); with best-fit Gaussians.}
\label{fig:Figure 9}
\end{figure}

The average gap size calculated analytically using the best-fit Gaussian and Equation 17, reveal $\Gamma_{ave}= 2.60 x 10^{-2} mm$, in agreement with the Monte Carlo simulation data.  

\subsection{Hula-Hoops and Angstroms }

Figure 10 represents the probability density as a function of gap size for a disk packing with disk radii $1.0 m \pm 10^{-10} m$.  This particular disk-packing was included in this study as it would represent an extreme physical minimum for $\Delta r$; that is, the characteristic scale for atomic radii.  This disk packing could be analogous to packing hula-hoops of radius $1.0 m$ with a very unrealistically tight mill tolerance of 1 Angstrom.  

\begin{figure}
\begin{center}
\leavevmode
\includegraphics[scale=0.5]{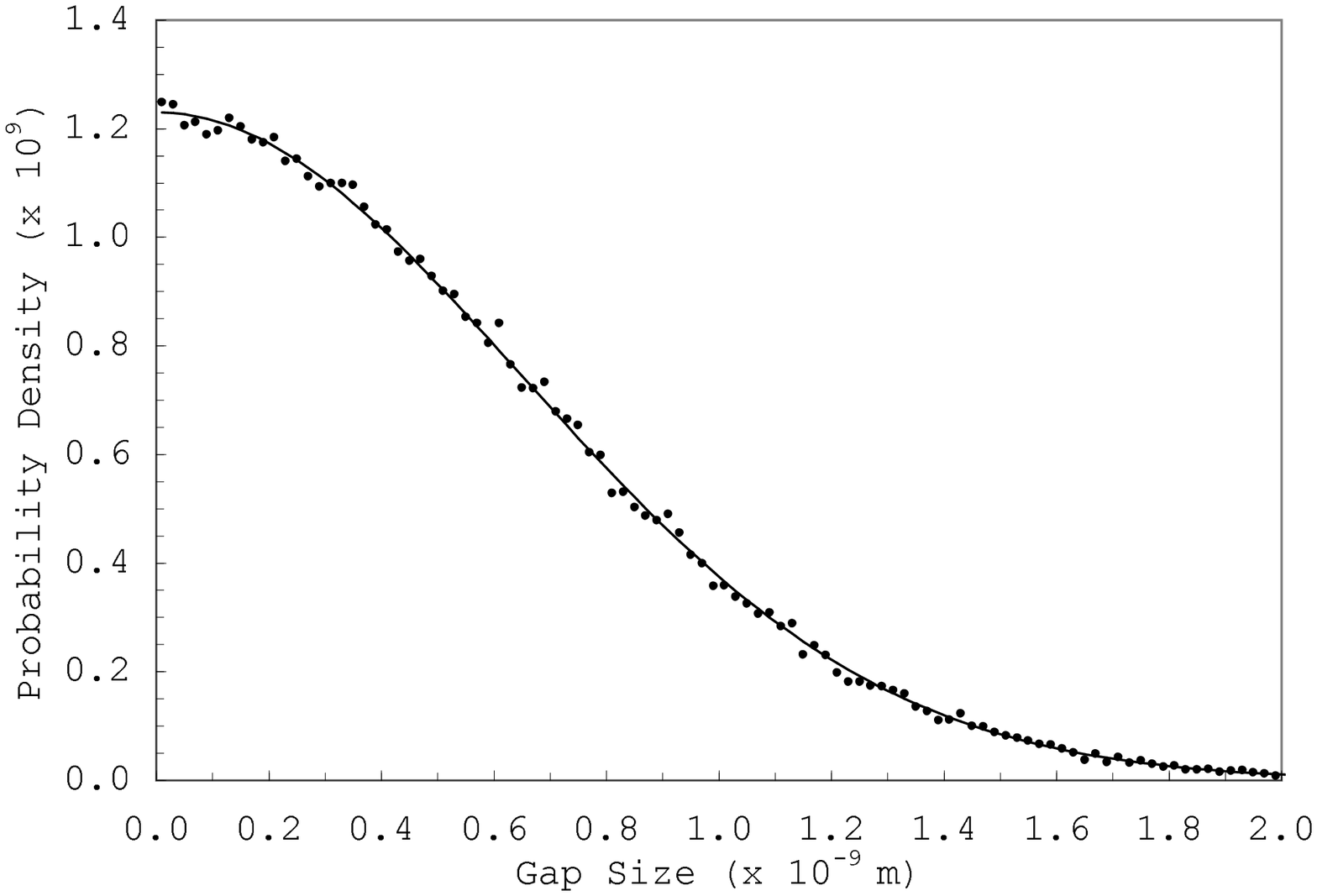} 
\end{center}
\caption{Probability density distribution for interparticle gap sizes arising in a packing of disks with $r = 1.0 m$ and $\Delta r = 1.0 x 10^{-10} m$, with best-fit Gaussian.}
\label{fig:Figure 10}
\end{figure}

As with the previous two studies discussed in Sections 7.1 and 7.2 above, the probability density distribution shown in Figure 10 is well represented with a Gaussian best-fit with the functional form of Equation 16.  The probability density data plotted and the best-fit curve both have standard deviation $\sigma = 6.49 x 10^{-10}$ m.  Figure 11 compares gap size probability density distribution arising using the angstrom specification $\Delta r = 1.0x10^{-10}$ m to those resulting when: i) $\Delta r$ is doubled to $2.0 x 10^{-10}$ m, and ii) $\Delta r$ is halved to $5.0 x 10^{-11} m$.  Doubling $\Delta r$ results in a broader distribution characterized with a standard deviation $\sigma = 1.30 x 10^{-9} m$, consistent with both simulation data and best-fit Gaussian.  Halving $\Delta r$ results in a sharper distribution characterized with a standard deviation $\sigma = 3.23 x 10^{-10} m$, consistent with both simulation data and best-fit Gaussian.  Consistent with the billiard and penny studies above, notice that doubling (halving) $\Delta r$ results in a corresponding doubling (halving) of the standard deviation, $\sigma$, the characteristic parameter defining the sharpness of the probability density distribution.

\begin{figure}
\begin{center}
\leavevmode
\includegraphics[scale=0.5]{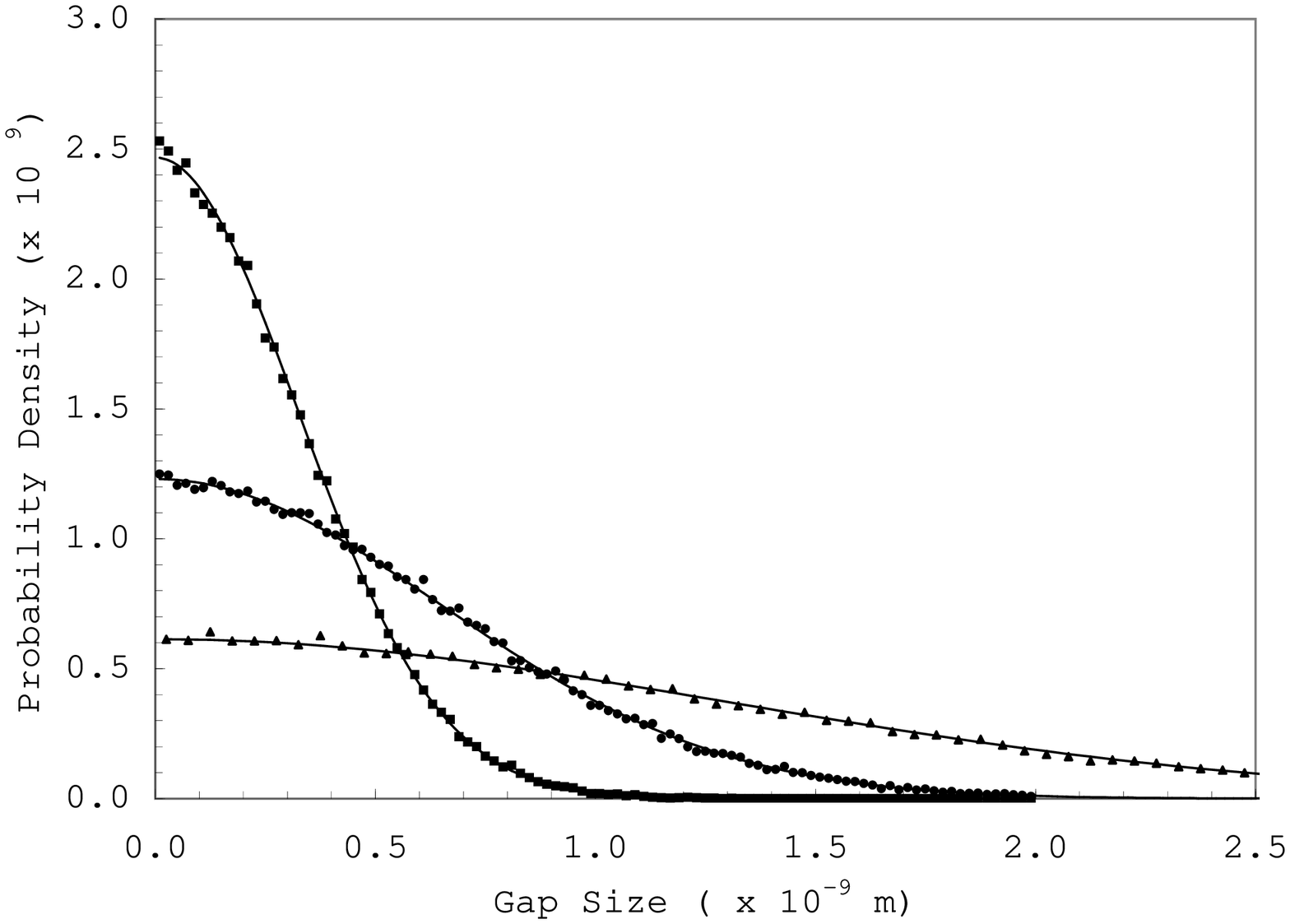} 
\end{center}
\caption{A comparison of probability density distributions for interparticle gap sizes arising in packings of disks with r = 1.0 m and i) $\Delta r = 2.0 x 10^{-10} m$ denoted by ($\blacktriangle$), ii) $\Delta r = 1.0 x 10^{-10} m$ denoted by ($\bullet$), and iii) $\Delta r = 5.0 x 10^{-11} m$ denoted by ($\blacksquare$); with best-fit Gaussians.}
\label{fig:Figure 11}
\end{figure}

The average gap size calculated analytically using the best-fit Gaussian and Equation 17, reveal $\Gamma_{ave}= 5.17 x 10^{-10} m$, in agreement with the Monte Carlo simulation data.  

\subsection{Universal Probability Density Distributions}

To allow comparison of the three very different ranges and corresponding disk packings detailed above, a normalized probability density distribution was generated as a function of normalized gap size, $\alpha = \frac{\Gamma}{\Delta r}$.  The best-fit Gaussian, shown on Figure 12, also uses a 'normalized' standard deviation $\sigma' = \sigma/{\Delta r}$ and is shown as the smooth curve.  The data for the three normalized probability density distributions exhibit exceptional agreement and all three of these probability density distributions are described by the same best-fit Gaussian with $\sigma = 6.43$.  This figure reveals the universal nature of the geometric frustration that arises in these disk-packings regardless of scale.  Figure 14 shows the qqplot for each of these normalized distributions comparing the distribution arising from the Monte Carlo simulation to the best-fit Gaussian.  The result for this case yields a correlation coefficient of $R=0.999991$ between the two distributions.  Each of the previous distributions yields a correlation coefficient at least as good when analyzing each distribution using similar qqplot analysis.

\begin{figure}
\begin{center}
\leavevmode
\includegraphics[scale=0.5]{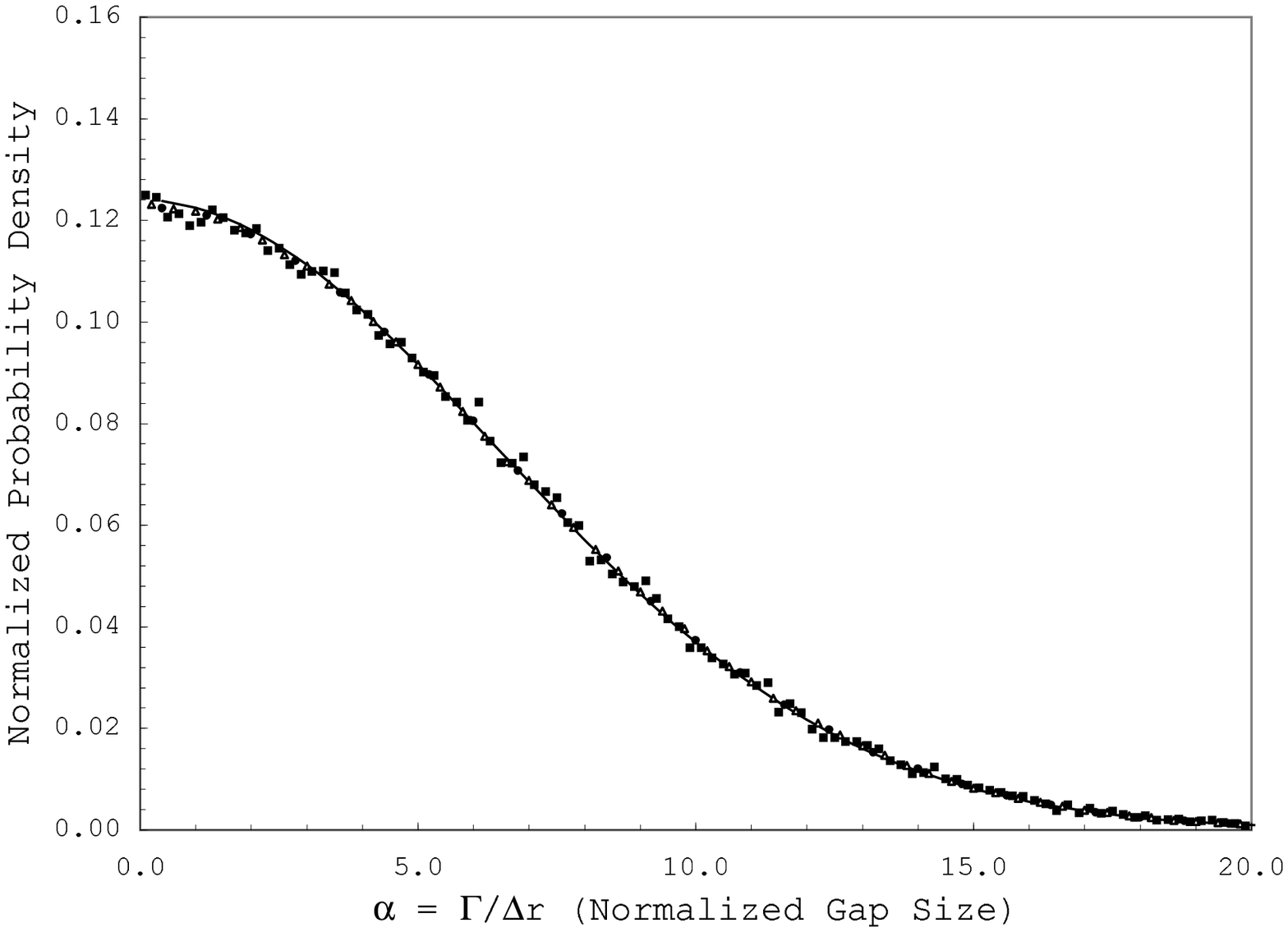} 
\end{center}
\caption{A comparison of normalized probability density distributions as a function of normalized gap sizes for i) billiard balls with BCA specs, ii) US Pennies with the specs of the US Mint, and iii) an extreme disk packing with $r = 1.0 m$ and $\Delta r = 1 x 10^{-10}$, with best-fit Gaussian.}
\label{fig:Figure 12}
\end{figure}

The average gap size calculated analytically using the best-fit Gaussian and Equation 17, reveal $\Gamma_{ave} = 5.09\alpha$, in agreement with the Monte Carlo simulation data.  

\subsection{Probability of a gap size $\Gamma \geq \alpha \Delta r$}

The probability of an interparticle gap arising, $\Gamma \geq \alpha \Delta r$, can be determined using standard probability theory, see Equation 18.

\begin{equation}
 P(\Gamma \geq \alpha \Delta r)=1-\int_{0}^{\alpha} \frac{2}{\sqrt{2\pi}\sigma} Exp[-\frac{\Gamma^2}{2\sigma^2}]d\Gamma 
\end{equation}

This integral has been evaluated for a range of $\alpha$, and the results are displayed in Figure 13.  The probabilities shown represent those for the universal probability distribution with $\sigma = 6.43$.  The reader should note for $\alpha \leq 5.0$, the probability of a gap arising $\Gamma \geq \alpha \Delta r$ is approximately $1-0.124\alpha$ as determined by Figure 13.  For increasing $\alpha$ (i.e. $\geq 5.0$) the probability becomes nonlinear as shown in Figure 13.
 
\begin{equation}
 P(\Gamma \geq \alpha \Delta r ) = 1-\int_{0}^{\alpha} \frac{2}{\sqrt{2\pi}\sigma} Exp[-\frac{\Gamma^2}{2\sigma^2}]d\Gamma  \sim 1-0.124\alpha \:for \: \alpha \leq 5.0
\end{equation}

\begin{figure}
\begin{center}
\leavevmode
\includegraphics[scale=0.5]{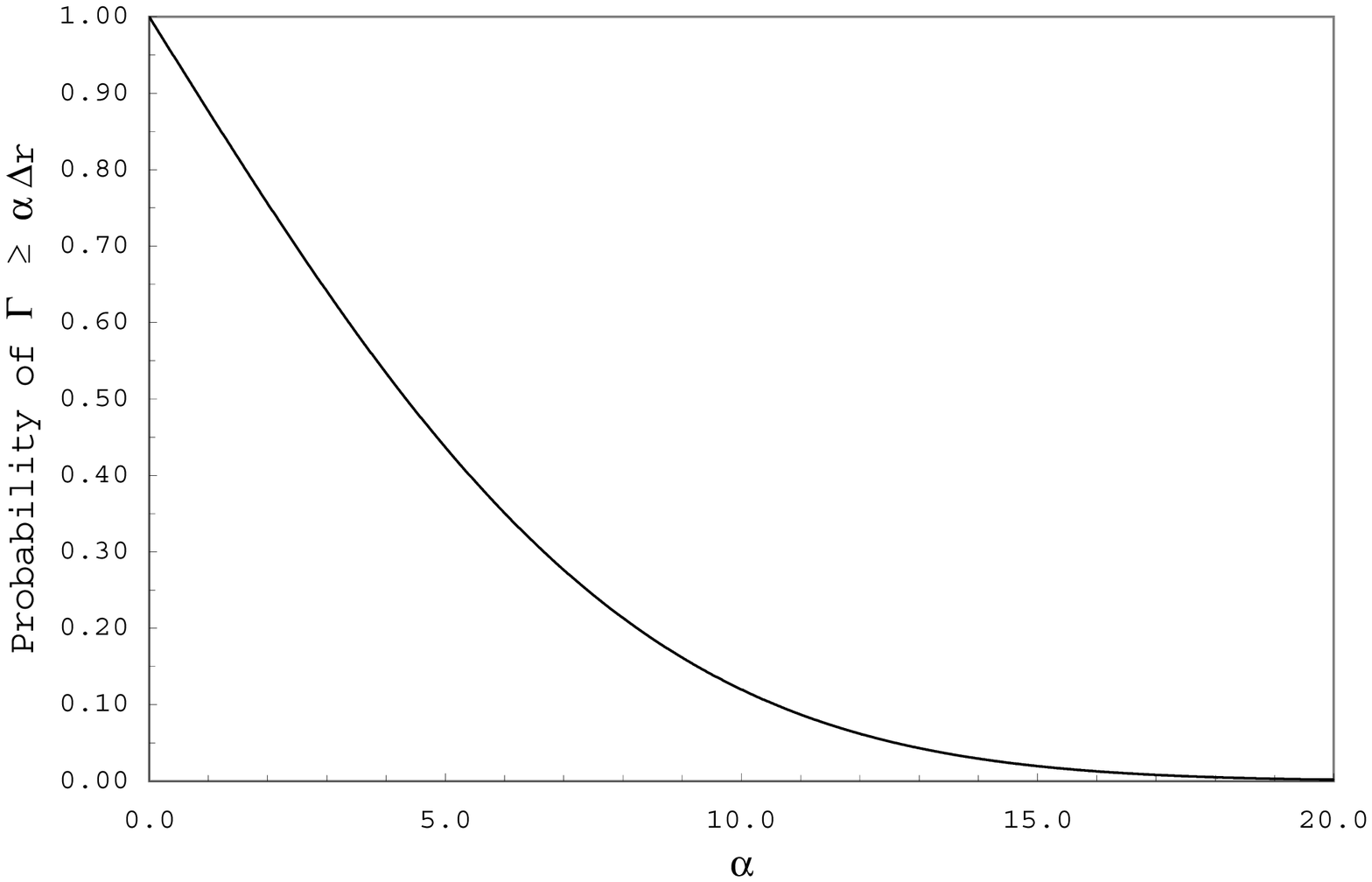} 
\end{center}
\caption{Probability of an interparticle gap arising with magnitude $\Gamma \geq \alpha \Delta r$, for a range of $\alpha$ with fixed $\Delta r$.  }
\label{fig:Figure 13}
\end{figure}

\begin{figure}
\begin{center}
\leavevmode
\includegraphics[scale=0.5]{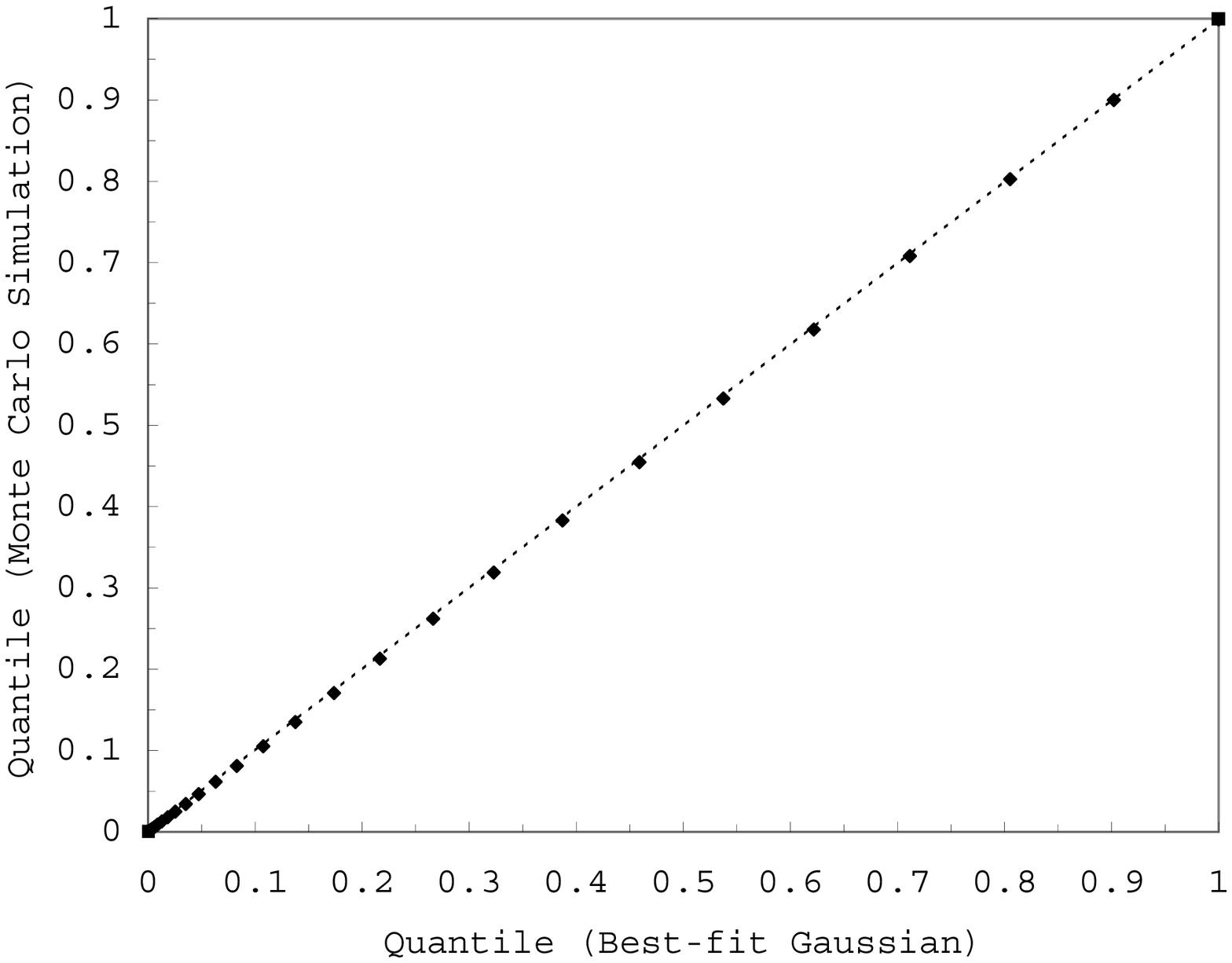} 
\end{center}
\caption{A qqplot comparing the actual distribution and the best-fit Gaussian for the case of normalized probability density distribution was generated as a function of normalized gap size.  Comparing the two distributions yields a correlation coefficient of $R=0.999991$.}
\label{fig:Figure 14}
\end{figure}

\section{Conclusions}

\subsection{General Comments}
This work analyzes the distribution and size of gap defects arising, locally, in an ensemble of non-interacting, hexagonal unit structures in the xy plane when packing disks with a Gaussian distribution of radii.  An analytic expression for the probability distribution and location of the gap defect of magnitude $\Gamma$ has been derived.  Due to the complexity of the expression a Monte Carlo simulation was employed to probe the resulting frequency and size distribution of gaps arising in an ensemble of $10^5$ unit structures.  A best-fit of the Monte-Carlo produced frequency distributions were performed and, in general, are represented very well with Gaussian best-fits as shown by a qqplot analysis for each case.

\subsection{Billiards and Granular Applications}

Any billiards aficionado realizes the importance of the break in any game of pool, especially nine-ball.  A novice will seek a re-rack to eliminate any gap that has suddenly appeared in a rack about to be scattered.  The experienced player realizes the opportunity the gap represents.  The presence of interparticle gaps drastically affect resultant scattering.  General trends can be detected for the initial direction of most of the balls in a nine-ball rack immediately after the break, primarily from experience.  However, an analytic expression accompanied with a probability distribution is the desired result.

The results of the present paper will be used in future studies as the propagation of shock waves through nine-ball billiard racks and hexagonal granular packs with interparticle gaps is considered.  More generally, upon the application of a force the scattering response of tightly packed granular materials and the affect of inevitable interparticle gaps upon this scattering will be considered.

Further, the results presented in this study were motivated by curiosities arising in billiard ball racks, however, the replacement of spheres with disks is a simplification to be removed in an upcoming study.  That is, an upcoming paper addresses gap defects arising in a packing of spheres, with a Gaussian distribution of radii, on the plane.

\subsection{Disk Packing Application}

The results of this work may also have application to Cowan type disk packings.  Whereas Cowan considered a random packing strategy for identical disks, the present study considered ordered packing of disks with random elements affecting the disk-size, not the packing strategy.  

Second, the probability distributions presented in this paper for interparticle gap size may find use in studies considering polydisperse disk packings covering the plane.  

In addition, although this work has not been an investigation of jamming in disk packs, the results may find application in this field since the interparticle gap size in these unit structures may directly affect the response of these packings upon the application of a disturbing force.

\subsection{Circle Packing Application}

The results of this work may have application to field of circle packing if one were to consider the possibility of an underlying pattern of tangencies with random elements affecting the length of each edge and location of each node.

\section{Acknowledgments}

The author extends thanks to Kenneth Stephenson for enlightening emails regarding circle-packings and Rhode Island College for extending faculty research release time in support of this work.


\end{document}